\input phyzzx
\nonstopmode
\sequentialequations
\tolerance=5000
\overfullrule=0pt
\nopubblock
\twelvepoint
\input epsf

\line{\hfill }
\line{\hfill PUPT 1594, IASSNS 95/111 }
\line{\hfill cond-mat/9602112}
\line{\hfill February 1996}

\titlepage

\title{Possible Electronic Structure of Domain Walls in Mott Insulators}

\author{Chetan Nayak\foot{Research supported in part by
a Fannie and John Hertz foundation
fellowship.~~~nayak@puhep1.princeton.edu}}

\vskip.2cm

\centerline{{\it Department of Physics}}
\centerline{{\it Joseph Henry Laboratories}}
\centerline{{\it Princeton University}}
\centerline{{\it Princeton, N.J. 08544}}

\author{Frank Wilczek\foot{Research supported in part by DOE grant
DE-FG02-90ER40542.~~~wilczek@sns.ias.edu}}

\vskip.2cm
\centerline{{\it School of Natural Sciences}}
\centerline{{\it Institute for Advanced Study}}
\centerline{{\it Olden Lane}}
\centerline{{\it Princeton, N.J. 08540}}
\endpage

\abstract{We discuss the quantum numbers of domain walls of minimal
length induced by doping Mott insulators,
carefully distinguishing between holon and hole walls.  We define a
minimal wall hypothesis that uniquely correlates the observed
spatial structure with the
doping level for the low-temperature
commensurate insulating state of La$_{2-x}$Ba$_x$CuO$_4$
and related materials at $x={1\over 8}$.   
We remark that interesting walls can be supported not only by
conventional antiferromagnetic but also by orbital antiferromagnetic
(staggered flux phase, $d$-density) bulk order.  We speculate on the
validity of the minimal wall hypothesis more generally, and argue that
it plausibly explains several of the most striking anomalous features
of the cuprate high-temperature superconductors.}

\endpage

\REF\ssh{W.P. Su, J.R. Schrieffer, and A.J. Heeger,
Phys. Rev. Lett. {\bf 42} 1698 (1979).}

\REF\polyarev{A. Heeger, S. Kivelson, J. R. Schrieffer, and W.-P. Su 
Rev. Mod. Phys. {\bf 60}, 781 (1988).}

\REF\jr{R. Jackiw, C. Rebbi
Phys. Rev. {\bf D13}, 3398 (1976).}

\REF\schulz{H.J. Schulz, J. Physique {\bf 50} 2833 (1989).}

\REF\tranquada{J.M. Tranquada {\it et al.}, Nature {\bf 375} 561 (1995).}

\REF\kanefisher{C.L. Kane and M.P.A. Fisher, Phys. Rev. Lett. {\bf 68}
1220 (1992); Phys. Rev. {\bf B 46} 15233 (1992).}

\REF\schulzgiam{T. Giamarchi and
H.J. Schulz, Phys. Rev. {\bf B 37}, 325 (1988).}

\REF\nakamura{Y. Nakamura and S. Uchida, Phys. Rev. {\bf B 46} 
5841 (1992).}

\REF\zaanen{J. Zaanen and O. Gunnarson, Phys. Rev. {\bf B 40}
7391 (1989).}

\REF\poilblanc{D. Poilblanc and T.M. Rice, Phys. Rev. {\bf B 39}
9749 (1989).}

\REF\emerykiv{V.J. Emery and S.A. Kivelson, Physica {\bf C 209}
597 (1993); Physica {\bf C 235-240} 189 (1994); in
Strongly Correlated Electronic Materials: The Los
Alamos Symposium 1993, eds. K.S. Bedell {\it et al.},
619, Addison-Wesley, Reading, MA 1994}

\REF\castroneto{A.H. Castro-Neto and D.W. Hone, Phys. Rev. Lett.
in press, cond-mat/9511079.}

\REF\fulco{T. L. Ho, J. Fulco, J. R. Schrieffer, and F. Wilczek
Phys. Rev. Lett. {\bf 52}, 1524 (1984).}

\REF\oaf{H.J. Schulz, Phys. Rev. {\bf B 39} 2940 (1989);
A.A. Nersesyan and A. Luther, unpublished;
A.A. Nersesyan, G.I. Japaridze, and I.G. Kimeridze,
J. Phys. Cond. Mat. {\bf 3} 3353 (1991);
A.A. Nersesyan and G.E. Vachnadze, J. Low Temp. Phys. {\bf 77} 
293 (1989);
L. Gorkov and A. Sokol, Phys. Rev. Lett. {\bf 69} 2586 (1992);
For a pedagogical review, see ``Spin-Singlet Ordering
Suggested by Repulsive Interactions,''
C. Nayak and F. Wilczek, cond-mat/9510132}

\REF\flux{I. Affleck and B. Marston, Phys. Rev. {\bf B 37}, 3774 (1988).}

\REF\jgfw{J. Goldstone and F. Wilczek,
Phys. Rev. Lett. {\bf 47} 986 (1981).}

\REF\wen{X.G. Wen, Phys. Rev. {\bf B 43} 11025 (1991);
Phys. Rev. Lett. {\bf 64} 2206 (1990).}

\REF\marshall{D.S. Marshall, {et al.}, ``Unconventional
Electronic Structure Evolution
with Hole Doping in Bi$_2$Sr$_2$CaCu$_2$O$_{8+\delta}$ 
Angle-resolved Photoemission Data'', Stanford preprint.}

\REF\baskaran{G. Baskaran, private communication.}

\REF\scsep{P. W. Anderson, G. Baskaran, Z. Zhou, and T. Hsu,
Phys. Rev. Lett. {\bf 58} 279 (1987).}

\REF\metals{C. Nayak and F. Wilczek, `` Physical Properties
of Metals from a Renormalization Group Standpoint,''
Intl. Jour. Mod. Phys. B. (in press),  cond-mat/9507040}

\REF\clarke{D.G. Clarke, S.P. Strong, and P.W. Anderson,
Phys. Rev. Lett. {\bf 72} 3218 (1994).}

\REF\sachdev{S. Sachdev, ``Quantum phase transitions in spins
systems and the high temperature limit of
continuum quantum field theories,'' Proceedings
of STATPHYS 19, cond-mat/9508080, and references therein.}

\REF\ando{Y. Ando and G.S. Boebinger, AT\&T Bell Laboratories
Preprint}

\REF\interlayer{J.M Wheatley, T.C. Hsu, P.W. Anderson,
Phys. Rev. {\bf B 37} 5897 (1988).}

\FIG\onedwalls{Holon and spinon walls for underlying antiferromagnetic
order in one dimension.  
Note that in each case there is a mismatch of the ordering at
the two ends, compared to the uniform ground state.  Simply removing
the possibility of occupancy at one site does not faithfully represent
the two degrees of freedom associated with an electron.  The spinon
configuration, which has two degenerate states but zero charge,
restores these.}

\FIG\twodwalls{Holon {\it versus\/} hole walls for underlying
antiferromagnetic order in two dimensions. The double-headed
arrows represent the possibility of choosing either spin direction.}

\FIG\altwalls{Alternative simple structures for oblique hole walls.}

It is familiar in several contexts that spatial defects, and
specifically domain walls, can have a profound effect upon electronic
behavior. This has perhaps been most thoroughly documented in the
case of polyacetylene [\ssh , \polyarev ], 
where (as we shall momentarily recall) it leads
directly to exotic quantum numbers for the elementary excitations.
Shortly after polyacetylene was analyzed, there was a resurgence of
interest in the classic Mott insulator problem -- spurred
by the discovery of the high-$T_c$ copper-oxide superconductors --
with particular emphasis on the nature of the low-lying excitations
which result upon doping. Several proposals revolve
around the idea that these excitations are solitonic.  
Recent experiments on La$_{2-x}$Ba$_x$CuO$_4$ and related materials
exhibit striking commensurability effects which, we shall argue,
acquire a simple but profound significance within this circle of
ideas.  Our analysis encourages us to consider generalizing these
ideas 
so as to  address the problem of anomalous behaviors in the 
copper oxides more generally, when commensurability fails.  On a
qualitative level, at least, the results appear quite encouraging.

\chapter{Quantum Numbers of Minimal Walls}

%interest in problem generally.  question of commensuration raised by
%1/8 phase
%recall poly-A analysis; pictures

Let us briefly recall the main results on quantum numbers of
domain wall defects in polyacetylene and related one dimensional
substances 
where a 
$Z_2$ symmetry breaking opens a gap at half filling.  They can be
modelled adequately, as regards quantum numbers,
using a simple scalar field $\phi$ to represent the order parameter,
and its   
interactions with the electrons by the Yukawa term 
$g \bar \psi \psi \phi$.  
The order parameter is supposed to take values 
$\langle \phi\rangle = \pm v$ 
in the uniform ground state, which induces a mass
term (gap) for the electrons.  If the mass is positive for one sign then
formally it is negative for the other; however viewed in itself
this sign is
meaningless because it can be reabsorbed into the definition of (one
chirality of) $\psi$.  However the change in sign at a domain wall is
significant, and leads to the existence of a normalizable
zero mode for $\psi$ localized at
the wall [\jr ].  This midgap state acquires half its spectral weight from
states in the original Dirac sea, and half from above.  Thus if it is
left unoccupied the state has electron number $-1/2$ while if it is
occupied the state has electron number $+1/2$.

Inclusion of the electron spin variable changes the picture.  There
are now two zero-modes, one for each spin.  If both are unoccupied,
the state is a spin singlet with electron number (and hence charge)
$-1$; if  both are occupied the electron number is $+1$, and of course
the electron number vanishes if one is occupied and the other empty.  
The charge $\pm 1$ states are spin singlets -- they represent the full
sea or its complement -- while the charge 0 states form a spin doublet.
Note that to reproduce the quantum numbers appropriate to
removing an electron -- a doped hole --
we require {\it two\/} domain walls, and must choose to take 
the charge $-1$ `holon'
state on one and the charge 0 doublet `spinon' one the other, {\it
i.e}. occupying exactly one of the four available zero-modes.

This analysis, which can be made completely formal and rigorous,
permits a simple pictorial representation (Figure 1).  It can also be
summarized in the
rubric $\{~~hole~~\rightarrow~~holon~~+~~spinon~~\}$,
with the understanding that both holon and spinon
are associated with wall defects.

%holon wall versus electron wall; electron wall fits commensuration

With this discussion in mind, it is easy to appreciate an important
distinction that arises for walls in two dimensions (Figure 2).  For
purposes of determining quantum numbers one can
consider a vertical two-dimensional wall as a collection of one-dimensional
walls -- one on each horizontal line -- and reduce to the preceding case.  
In this way we see that the vacant wall, considered in
depth by Schulz [\schulz ],  does not correspond directly
to the quantum numbers one would associate with an ordinary hole;
rather it gives an array of holons.  To represent the
degrees of freedom of weakly interacting holes faithfully, 
one must instead allow
the more complicated configuration shown on the right.  Note that this
configuration corresponds to $1\over 4$ filling along the wall.
Indeed, this is basically the
same factor $1\over 4$ we encountered two paragraphs ago.  
Note that there is nothing obviously special about this filling,
viewed in itself; its distinction resides only in its pedigree. 

The pictorial representation in Figure 2 makes the conclusion of the
formal analysis quite appealing
intuitively.  The idealized
hole wall schematically indicated there contains equal numbers of 
two kinds of sites, empty and singly occupied.  The empty
sites carry charge (relative to the uniform bulk state) but no spin:
they carry holon quantum numbers.  The singly occupied sites have a
spin degree of freedom each of whose orientations is equally
favorable, due to an evident symmetry, but no deviation from the
uniform bulk density.  They carry spinon quantum numbers.  Together,
each such pair has the quantum numbers of a single hole.  When one
introduces interactions and hopping to this static picture the
low-energy eigenstates will be quite different in structure, but one
would not expect the    
quantum numbers to be altered.

It is remarkable that the quantum numbers of hole ({\it not\/} holon) 
walls are just what is needed to
reproduce
the spatial structure revealed in neutron scattering  
by Tranquada {\it et al}. for the low-temperature insulating state
of La$_{1.6-x}$Nd$_{0.4}$Sr$_x$CuO$_4$ at the appropriate 
commensurate doping $x = {1\over 8}$  
[\tranquada ].
The neutron scattering  results indicate
stripes modulating the bulk spatial structure
along every fourth row of atoms, and this will occur precisely at 
$x = {1\over 8}$ when the walls are, in the sense we have defined above,
hole domain walls.

One could of course imagine adding extra domain wall bits not
associated with doping, or
conversely hole excitations which are not attached to domain walls.
However the narrowness of the observed non-superconducting phase near
$x= {1\over 8}$ implies and is implied by the minimal wall hypothesis:
that exactly
enough
domain wall
is produced, at least at low temperatures, to provide a midgap
home for the dopants.

To avoid possible misunderstanding, let us emphasize that we do not
propose it as a general law that the hole quantum numbers must
be represented faithfully.  One can certainly imagine that the spins
of the dopants are frozen, as in the underlying bulk state, or that
the charges are frozen.  That is, there could be phases where 
the quantum numbers of either holon or spinon domain walls alone are
realized -- indeed, it appears [\tranquada ] that in La$_2$NiO$_2$
the former possibility occurs.  Rather we
are proposing, much in the spirit of Landau's Fermi liquid theory,
that there is a universality class corresponding to faithful
representation of the weakly-coupled dopant degrees of freedom, and
that the minimal realization of this class occurs in the situation mentioned.
Presumably the hole domain wall is favored when 
Coulomb repulsion dominates the antiferromagnetic
correlation energy and the holon wall is favored
in the opposite case. 

Our picture of the $1\over 8$ state has significant dynamical
consequences.  According to it, the charge carriers
in this state are contained in an
infinite array of (coupled) 1-dimensional metals.
These degrees of freedom should dominate the specific heat
and this should be proportional to $T$.
Most interesting and characteristic, however, is
the resistivity. Impurities in one-dimensional metals (Luttinger
liquids) with repulsive interactions 
lead to a (non-universal) power-law divergence of the resistivity
[\kanefisher , \schulzgiam ].
By contrast, weak Anderson localization in
two dimensions leads to a logarithmic
divergence. In the La$_{1.6-x}$Nd$_{0.4}$Sr$_x$CuO$_4$
resistivity data of Nakamura and Uchida [\nakamura],
one sees a slight positive curvature in the in-plane resistivity
at temperatures above the structural phase transition
which presumably pins the domain walls. In La$_{2-x}$Sr$_x$CuO$_4$,
this positive curvature is the precursor of the
logarithmic divergence of weak localization.
Below the structural transition, however, there is a
striking increase in the divergence of the
resistivity. It is attractive to interpret this
marked increase as resulting from the effects
of impurities on the essentially one-dimensional
transport which takes place in pinned domain walls.

Some other ideas relating to the existence
and significance of domain walls
and antiferromagnetic stripes may be found in
[\zaanen-\castroneto].

\chapter{Analysis of Single Walls}

Let us very briefly indicate the shape a more mathematical treatment
of the preceding would take. The problem of domain
walls in a doped antiferromagnet was discussed by
Schulz [\schulz]; the reader should compare [\fulco ] where
a formally complete but highly compressed discussion of a closely
related problem is given.  Consider the Hubbard model on a two-dimensional
square lattice:
$$
H ~=~ 
-t\sum_{\rm n.n. \langle ij\rangle} c^\dagger_{\alpha, i}\,
c^\alpha_i 
~+~ U \sum_{\rm i} {n_i^\uparrow}\,{n_i^\downarrow}
\eqn\hubmodel
$$
where $\alpha$ is a spin
index. 
In the spin-density wave formalism, we assume an
antiferromagnetic background with $S_z = (-1)^{m+n} S$
($i=(m,n)$).
A domain wall solution has $S_z = (-1)^{m+n} S$ for
$n\geq 1$ and $S_z = (-1)^{(m+n+1)}  S$ for $n\leq-1$.
In the presence of this mean field, the factorized
form of the Hamiltonian \hubmodel\ is:
$$
H ~=~ 
-t\sum_{\rm n.n. \langle ij\rangle} c^\dagger_{\alpha, i}\,
c^\alpha_i 
~+~ U S \sum_{\rm i} \sigma(n){(-1)^i}\,({n_i^\uparrow}-{n_i^\downarrow})
\eqn\fachubmodel
$$
where $\sigma(n)=+1$ for $n\geq 1$, $\sigma(n)=-1$
for $n\leq-1$, and $\sigma(0)=0$. This single-particle
Hamiltonian may be diagonalized and, as usual,
one finds conduction and valence bands split
by a gap; the self-consistency condition that
the ground state of the Hamiltonian \fachubmodel\
exhibit the required antiferromagnetic background
leads to a BCS-like gap equation.
For our purpose we want to focus on the essentially new states which
arise in the presence of the domain wall.
Physically, it is not hard to see why such states should
exist: an electron located at the wall essentially
sits in a potential well in the direction perpendicular
to the wall but is free to move along the wall.
At the level of the single-particle Hamiltonian
\fachubmodel, these states are of the form:
$$
\psi_k (m, n) ~=~  e^{ik ma} \Big( {e^{i(k+\pi)na}} +
i {(-1)^{m+n}} {e^{i(k+\pi)na}} \Bigr)
{e^{-{\kappa_k} |na|}}
\eqn\modeform
$$
with 
$$
 \sinh ({\kappa_k} a ) ~=~ {{US}\over{2t \sin ka}}
\eqn\kappaeqn
$$
and energy
$$E = 2t \cos pa (\cosh {\kappa_k}a - 1) \eqn\energyeqn$$
In the small-$U$ limit, the domain-wall band
becomes narrow, as was found by Schulz [\schulz].
A more sophisticated treatment must derive
the form of the factorized
Hamiltonian self-consistently.

Such formal arguments can be made to seem more powerful than they are,
since they are intrinsically weak-coupling arguments and involve 
extrapolating to reach any realistic (or even finite) coupling. In
calculating
digital
quantities such as the number of states with specified 
quantum numbers extrapolation of this sort 
has a fair chance of success, but
even here real physical questions may arise.  For example, when we
consider oblique hole
domain walls two structures naturally suggest themselves, as depicted
in Figure 3.  The density of holes per unit length of domain wall is
twice as great for one as for the other; one or the other, even a superposition
might be most favorable, depending on detailed energetics.

\chapter{Alternative Bulk Orders}

In the preceding section we have framed the discussion in the context
of antiferromagnetic bulk ordering.  It is important to emphasize,
however, that the underlying principles leading to the quantum number
and commensurability assignments are more general.  
In particular, they apply to a different kind of
order known in its two-dimensional incarnation
variously as orbital antiferromagnetism, staggered flux,
or d-density [\oaf,\flux].

We will use the last name, since the others
are inappropriate for our extrapolations outside two dimensions.
The issue of other types of ordering is not merely academic.
The neutron scattering data of [\tranquada] indicate
that there is first a transition to a striped phase
without spin ordering -- perhaps consisting of domain walls
in the alternate ordering that we consider below -- and then a
transition
to a phase with antiferromagnetic stripes.

For later purposes
   it is important to define d-density order in a form that applies to other
   dimensionalities (especially 1!), as follows.  Let the lattice spacing
   be $a$ in all directions.  Then we suppose the existence of a
   non-trivial
   correlation
   $$
   \langle c^\dagger_\alpha (k) c^\beta (k+Q)\rangle 
~=~ i\,\delta^\beta_\alpha f(k)
   \eqn\corr
   $$
   where $Q$ has components of magnitude $\pi/a$ in all directions and
   $f$ is a real function that changes sign upon $\pi/2$ rotation around
   any axis.  This correlation breaks symmetry under time reversal $T$,
   translation by one lattice spacing, and $\pi/2$ rotations, but in such
   a way that the square of any of these operations, or the product of
   any two, is a valid symmetry.  This order, although its direct
   manifestations in neutron scattering (for example) are quite subtle,
   effectively halves the size of the Brillouin zone and naturally
   induces Mott insulator behavior at half filling.  In two dimensions
   the energetics for this kind of ordering appear quite favorable,
   especially in view of the nesting property.  Moreover since the
   symmetry breaking is $Z_2$ one very naturally has topologically stable
   domain walls in this case, and they support zero modes as analyzed
   above.

   %ordering known by various names; symmetry properties

At and near half-filling, such a state may be thought of as a $d$-wave
spin-singlet particle-hole paired state, just as a
charge-density-wave is an $s$-wave particle-hole paired state.
The $i$ on the right-hand-side of \corr\ is necessitated
by the $d$-wave symmetry of $f(k)$. A BCS-like paired wavefunction
may be written down for a state with such ordering:
$$
\Psi ~=~ \prod_k \, (u_{k\uparrow} c^\dagger_{k\uparrow}
+ v_{k\uparrow} c^\dagger_{k+Q \uparrow} )\,
(u_{k\downarrow} c^\dagger_{k\downarrow}
+ v_{k\downarrow} c^\dagger_{k+Q \downarrow} )\, |0 > ~,
\eqn\bcswf
$$
where the product runs over some subset of the interior of the
magnetic zone (i.e. the diamond $(k_x \pm k_y )a = \pm \pi$),
and $|u|^2 + |v|^2 =1$ for all
values of the indices, to preserve normalization, and the number of
factors is determined by the density of electrons (see below).
The product ${u_k}{v_k}$ is imaginary; its magnitude is proportional
to the gap.
In two dimensions, $d$-density state exhibits staggered flux. In the case of
the ansatz wavefunction \bcswf,
$$\langle\,{c^{\dagger\alpha}_{\bf x+a}}\,{c_{{\bf x}\alpha}}\,\rangle
= {\sum_k}\,{e^{-i{\bf k\cdot a}}}\,\Bigl({{\bar u}_{k\alpha}}\,{u_{k\alpha}}
-{{\bar v}_{k\alpha}}\,{v_{k\alpha}}\Bigr)\,\,+\,\,
{e^{-i{\bf Q\cdot x}}}\,\,{\sum_k}\,{e^{-i{\bf k\cot a}}}\,
\Bigl({{\bar u}_{k\alpha}}\,{v_{k\alpha}}
-{{\bar v}_{k\alpha}}\,{u_{k\alpha}}\Bigr)\eqn\staggflux$$
If the $u_k$'s and $v_k$'s are time-reversal invariant,
${u_k}={u_{-k}}$, ${v_k}={v_{-k}}$, then the first 
sum is purely real while the second is purely imaginary
for $d$-density ordering. The flux through a plaquette
with its upper left corner at ${\bf x}$ is the fourth
power of \staggflux; the flux through a neighboring
plaquette is the complex conjugate. Since the $d$-density
state exhibits staggered flux, it is presumably in the same
universality class as the staggered flux state. However, the variational
state \bcswf\ differs from the staggered flux variational
states in that the real part of the hopping matrix
element is kept fixed and the imaginary part varied
in the former case while the magnitude is kept fixed
and the phase varied in the latter.

   %mean field and fluctuations; energetic advantage of 2 dimensions
   %and nesting

   Because the ordering in question breaks only a discrete symmetry, it
   becomes particularly favorable -- as compared to potential rival 
   orders breaking a continuous symmetry -- in low dimensions, due to its
   relative immunity from fluctuations.  A mean-field analysis of
   extended
   Hubbard models with nearest-neighbor repulsion
in 2+1 dimensions indicates the favorability of this
   ordering for a range of parameters.  But  perhaps the most
   profound and compelling argument for its physical significance comes
   from analysis of the 1+1 dimensional Hubbard model near half filling,
   as we now sketch.

   The Fermi surface splits into two points, describing left- and
   right-movers.
   Spin and charge excitations travel at different velocities, and are
   conveniently represented by different fields.  Let the charge field
   for left and right movers be represented by the appropriate
   components $\chi_{L, R}$ of a scalar
   field $\chi$ compactified on a circle of radius $\sqrt 2$.  At generic
   values
   of the filling, these are free fields.  Precisely
   at half filling there is an additional relevant interaction, the 
   Umklapp process, that in
   terms of the electron fields is
   $$
   {\cal L}_{\rm umklapp} ~=~ 
   u(\epsilon_{\alpha\beta} \psi^{\dagger\alpha}_L \psi^{\dagger\beta}_L )
    (\epsilon^{\gamma\delta} \psi_{R\gamma} \psi_{R\delta} )
   \eqn\electumk
   $$
   and for the holon fields induces the simple form
   $$
   {\cal L} ~=~ {1\over 2} (\partial\chi)^2 ~+~ 2u \cos \sqrt 2 \chi~.
   \eqn\holonumk
   $$ 
   Here $\chi \equiv \chi_L + \chi_R$ and $u = {U\over 16t}$, where $t$ is
   the
   hopping parameter.  
   For an attractive interaction $u < 0$ the energy is minimized
   at
   $\langle \chi \rangle = 0 ~{\rm or}~\sqrt 2 \pi$, 
   and for a repulsive interaction $u > 0$ the
   energy is minimized at $\langle \chi \rangle = \pm \pi/\sqrt 2 $.

   In terms of the holon variables
   $\psi_L \equiv e^{-i\chi/\sqrt 2}$, 
   $\psi_R \equiv e^{i\chi_R/\sqrt 2}$, $\langle \chi \rangle = 0 $
   corresponds
   to 
   $
   \langle \psi_{R} \psi^{\dagger}_L \rangle ~=~
   \langle \psi_{L} \psi^{\dagger}_R \rangle ~=~ c~,
   $
   a real number.  This is a conventional, commensurate charge density
   wave.  On the other hand $\langle \chi \rangle = \pi/\sqrt 2$
   corresponds to
   $$
   \langle \psi_{R} \psi^{\dagger}_L \rangle ~=~
   -\langle \psi_{L} \psi^{\dagger}_R \rangle ~=~ if ~,
   \eqn\onedeorder
   $$
   a pure imaginary number.  As a result of the minus sign this state
   does not have charge-density order, contributions from L-R and 
   R-L terms cancelling.  Instead, \onedeorder\ is precisely of the
   d-density form.

   The one-dimensional model suggests another interesting although more
   speculative possibility.  If we assume that spin-charge separation
   occurs in two dimensions as it does in one, then we can imagine that
   the holon degrees of freedom undergo d-density ordering, opening a
   gap for charged excitations but leaving gapless spin excitations.

   In any case, in the one dimensional model the charged excitations are
   associated with domain walls.  Indeed one has the relationship 
   $j_0 = {1\over 2\pi \sqrt 2} \partial_1 \chi$, indicating a unique
   connection between domain walls and localized charge 1/2 inhomogeneities.  

Since $d$-density ordering breaks a $Z_2$ symmetry, domain
walls are topologically stable. As we will see below, we
can use more powerful arguments than those of section 2
to show that these walls support gapless modes. In order to see that
they do, we can calculate the currents that are built up
by an external electromagnetic field
in the presence of a domain wall, using the method of Goldstone
and Wilczek [\jgfw]. Equivalently,
we can calculate the terms which are generated in the
low-energy effective action for the electromagnetic field
when the gapped fermionic excitations are integrated out. 
This action will fail to be gauge-invariant,
necessitating the existence of gapless excitations at the
domain walls which restore gauge invariance as in the
case of the edge excitations in the quantum Hall effect [\wen].

The result of such a calculation is the effective
action:

$$\eqalign{{S_{\rm staggered C.-S.}}\,\,=&\,\,\int {d^3}k \,\,
{\epsilon_{\mu\nu\lambda}}\,\, {F_{\nu\lambda}}({\bf Q-k})\,\,
\biggl((\cos{k_x}a + \cos{k_y}a - 2)\,{A_\mu}({\bf k})\cr
&\,\,+\,\,
i{Q_\mu}\Bigl((\cos{k_x}a - 1) {A_x}({\bf k})\,
+\,(\cos{k_y}a - 1) {A_y}({\bf k})\Bigr)\biggr).\cr}
\eqn\scskspace$$
or, in real space:
$${S_{\rm staggered\, C.-S.}}\,=\,\int{d^3}x\,\,{\epsilon_{\mu\nu\lambda}}\,
\,{e^{i{\bf Q\cdot x}}}\,\,{\eta_\mu} 
\Bigl({A_\mu}({\bf x}+a{\bf \hat\mu})
- {A_\mu}({\bf x})\, +\, i{Q_\mu}{\int_x^{x+a{\hat\mu}}}A \Bigr)
\,\,{F_{\nu\lambda}}({\bf x})\eqn\staggcs$$

To see that there must be gapless excitations at a domain
wall, suppose to the contrary 
that there are no low-energy excitations
other than photons. This  is inconsistent because
the action \staggcs\ is not gauge invariant. Under a gauge
transformation, ${A_\mu}\rightarrow{A_\mu}+{\partial_\mu}f$,
the action transforms as:
$$\eqalign{\delta S &= \,\int {d^3}x \,\,{\epsilon_{\mu\nu\lambda}}\,
{e^{i{\bf Q\cdot x}}}\,\,{\eta_\mu}
\biggl({\partial_\mu}f({\bf x}+a{\bf\hat\mu })
- {\partial_\mu}f({\bf x})\,
\,+\,i{Q_\mu}\Bigl(f({\bf x}+a{\bf\hat\mu})-f({\bf x})\Bigr)\biggr)
\,\,{F_{\nu\lambda}}({\bf x})\cr
&=\,\int {d^3}x \,\,\,{\epsilon_{\mu\nu\lambda}}\,\,\,{\partial_\mu}
\Biggl(\,{\eta_\mu}\,{e^{i{\bf Q\cdot x}}}\,\,
\Bigl(f({\bf x}+a{\bf\hat\mu})
-f({\bf x})\Bigr)\,\,
{F_{\nu\lambda}}({\bf x})\Biggr).\cr}\eqn\staggcsgt$$
Thus, the action fails to be gauge-invariant at the
boundary of the $d$-density ordered region, i.e. at
a domain wall. This is rectified, however, if there
are gapless modes at the domain wall with the action
$${S_{\rm domain\,wall}}\,=\,\int {d^2}x\,\, {({D_\mu}\varphi)^2}\,\,
+\,\,{{\hat n}_\mu}{\eta_\mu}\,\,{e^{i{\bf Q\cdot x}}}\,\,
\Bigl(\varphi({\bf x}+a{\bf\hat\mu})-\varphi({\bf x})\Bigr)\,
{\epsilon_{\mu\nu}}{F_{\mu\nu}}({\bf x})
\eqn\wallaction$$
where $D_\mu$ is the ordinary covariant derivative,
and $\varphi$ transforms as $\varphi\rightarrow\varphi+f$
Since the latter term can be rewritten as
${{\hat n}_\mu}{\eta_\mu}\,{e^{i{\bf Q\cdot x}}}\,
\Bigl({F_{\mu\nu}}({\bf x}+a{\bf\hat\mu})-{F_{\mu\nu}}({\bf x})\Bigr)\,
{\epsilon_{\mu\nu}}\varphi({\bf x})$, it is clear that this
theory describes an ordinary, non-chiral scalar
field so long as the electromagnetic fields are
slowly varying on distances of order the lattice spacing.

\bigskip

\chapter{Possible Dynamical Implications}

Given our interpretation of the state at $x={1\over 8}$, it is natural
to speculate that the minimal domain wall hypothesis is valid more
generally -- that the physical effect unique to this doping lies
not in 
the nucleation of hole walls to accommodate holes, but only in that the
resulting walls get locked into a rigid spatial structure.

Some recent photoemission data [\marshall] on
Bi$_2$Sr$_2$CaCu$_2$O$_{8+\delta}$
is most suggestive in this regard [\baskaran].  The
data suggests the existence of  significant flat regions of the Fermi
surface at or near $k_x = \pm {\pi \over 4a}$ or  $k_y = \pm {\pi
\over 4a}$.  But these are precisely what one expects from horizontal
and vertical hole walls!  Indeed, for a horizontal wall $k_y$ is indefinite
while for the $1\over 4$ filling characteristic of hole walls
the Fermi points occur at $k_x = \pm {\pi \over 4a}$.  
Several potential complications should be noted. 
The flat regions occur do not extend to the smallest values of $k_y$,
but that is where one is sampling over a large geometric region, and
the possibility for interwall interactions exists.  Also, there could well
be low-energy electronic excitations in the bulk state, as occurs for
special points on the Fermi surface near half-filling in the d-density
and flux phases at $(\pm {\pi\over 2a}, \pm {\pi \over 2a})$.  Also, at
large $k$ the form-factor of the zero modes will cut down their
response
from that expected for
point-like particles.  The appearance and strength of the indicated
flat regions, then, requires
(on our interpretation) specific properties of the horizontal and
vertical walls -- in particular, if the flat regions extend to large
momenta, one requires
that these walls are effectively quite narrow in the transverse direction.

The
dynamics of interacting domain walls with non-trivial electronic
structure appears to be a new and challenging problem in the context
of condensed matter physics; it includes a form of finite-density
string theory.  The discussion
that follows represents what we think are plausible conjectures, but
is very far from rigorous.

%spin-charge separation and 1d dynamics (strong ee interactions)

An first qualitative consequence of these ideas is that the
behavior of the electronic degrees of freedom, being described by a 1+1 
dimensional theory, are much more affected by interactions,
even 
at low
temperature, than in a homogeneous phase.  
One must start their description 
with the Luttinger liquid theory,
including spin-charge separation, rather than the Fermi liquid theory.
Anderson has championed this point of departure for some time, at
least partly on phenomenological grounds [\scsep ].

A novel  source of electrical resistance in
domain wall transport is the interaction of 
gapless charged excitations which carry currents
with transverse vibrations of the domain
wall.  The natural effective theory
for the fermionic excitations interacting with
these long-wavelength `phonon' excitations of the wall is:
$${\cal L}\,\,=\,\,{{\cal L}_{\rm fermion}} + 
{\bigl({\partial_t}\phi\bigr)^2}-{c^2}{\bigl({\partial_x}\phi\bigr)^2}
\,\,+ \,\,\lambda\,{\psi^\dagger}\psi\,{\partial_x}\phi\eqn\llphtheory$$
where $\phi$ is the domain-wall `phonon' displacement
field. We can diagonalize the bosonized form of this theory. It is
straightforward to show that the conductivity yielded by
such a calculation is of the same form as for a non-interacting
electron system, $\sigma(\omega)\propto\delta(\omega)$. This
runs counter to one's 
intuitive expectation that currents can be dissipated
when charge carriers emit phonons. The problem, of course,
is that unless the `phonons' interact with 
other degrees of freedom they return all of
their energy and momentum to the charged currents.  In our case we expect the
phonons to be quite strongly coupled to the bulk degrees of freedom,
and
to lose their energy before they can
return it to the electrons, so
$$\sigma\sim {1\over T}.\eqn\condscal$$
As we have recently emphasized, 
this is the naive scaling for temperature dependence of conductivity
in a theory with
a Fermi surface [\metals ]. 
In a conventional Fermi liquid it is usually modified 
because the resistance is due to so-called dangerous irrelevant
operators, which introduce additional powers of $T$.
In \llphtheory, all interactions are marginal, so we expect
the naive scaling \condscal\ to hold.

Another issue of great importance is the sensitivity of wall
conductivity to impurities.  The leading interaction introduced by an
impurity is of the type
$$
\Delta {\cal L} ~=~ \delta^1(x) \psi^\dagger \psi~,
\eqn\impure
$$
and is marginal by power counting.   For repulsive electron-electron
interactions this operator becomes relevant, and one expects large
localization effects, as we have already mentioned in the context of
the
$x={1\over 8}$ state.  However when the domain walls are not pinned,
exchange of 
the wall vibration `phonons', as well as of ordinary phonons,
generates a competing attractive interaction.  If the attractive
interaction wins out, then the impurity interaction is {\it
irrelevant} [\kanefisher].  
For a random distribution of
impurities, the same is true if the attractive interaction
is sufficiently strong [\schulzgiam].
Attraction in the Cooper channel, which in higher
dimensions triggers an instability toward superconductivity, cannot do
so in 1 dimension because of large fluctuation effects.
It leads instead to unusually good (but not quite super) conductivity,
insensitive to impurities. The divergent resistivity seen
in the experiments referred to earlier is
presumably the result of the fact that the pinning
of the domain walls by a structural phase transition
pushes the `phonon' excitations to high energies. As
a result, the attractive interaction between electrons
is reduced and the repulsive case -- with 
markedly increased sensitivity to
impurities --
applies.

Suppose, now, that there are
a number of domain walls in the system. If these walls are
close enough charge can freely tunnel from one to another.
This is a relevant perturbation. Clarke,
Strong, and Anderson [\clarke ] have argued that 
tunneling between one-dimensional
Luttinger liquids may be an incoherent process.  Then
the tunneling conductivity vanishes as $T\rightarrow 0$.
Even if their arguments were incorrect, the tunneling
between domain walls will be incoherent if the walls
are not parallel or are fluctuating since, in
such cases, the momentum parallel to the walls cannot be conserved.
There are then two possibilities for the resistance between
two points in a sample. If the two points are continuously
connected by domain walls, then the resistance is
simply linear in $T$. If the two points are not continuously
connected by domain walls, then some inter-wall tunneling
is necessary for charge to be transported from one point
to the other. In such a case, the resistance is linear
in $T$ at high temperatures, where the inter-wall tunneling
varies slowly with the temperature and is not yet the limiting factor.
At low temperatures, the resistance should turn up
and diverge as $T\rightarrow 0$.

As the doping is increased from zero,
we expect domain walls to proliferate until, finally,
a percolation critical point is reached. If the doping
is less than this critical value, then we expect the sample
to be metallic at high temperatures with $\rho\sim T$
as discussed above and then to cross over to insulating behavior
at low temperatures. At the percolation threshold,
we expect metallic behavior, with $\rho=aT$ extrapolating
to $\rho=0$ at $T=0$ due to the insensitivity to impurities noted
above. 
The underlying bulk order is
destroyed at some doping level at or above the percolation
critical point. When this occurs, Fermi liquid behavior
is expected. This state of affairs may be summarized by
saying that exotic metallic behavior is expected in
the ``quantum critical regime'' [\sachdev] centered
at the percolation critical point.

At any doping level less than that which restores
Fermi liquid behavior,
$c$-axis transport, i.e. transport out of the plane, should be
qualitatively similar to in-plane transport in the underdoped
regime. At high temperatures, $\rho\sim T$ is expected,
but at low temperatures, the incoherence of tunneling between
domain walls in neighboring planes causes insulating
behavior.

All this is strikingly reminiscent of behavior recently observed 
experimentally in [\ando ], if we suppose the percolation
threshold is close to the optimal doping level  
for superconductivity.  The correlation of maximal 
$T_c$ with a bifurcation in the qualitative behavior of 
the low-temperature normal state resistivity is a striking `coincidence', 
which as far as
we know has not previously been illuminated theoretically. 
It can be motivated, within the circle of ideas we have been
advocating,  by a plausible
extension of the interlayer tunneling hypothesis [\interlayer], which
has had considerable semiquantitative success, to intralayer,
interwall
tunneling.  Indeed, supposing that the transition to
superconductivity is triggered even at the planar level by the gain in
delocalization energy when pairs can propagate coherently over the
plane, it seems eminently reasonable that the largest gain in accessible area
per unit strength of pairing takes place near the percolation
threshold, where the accessibility of large areas hangs in delicate balance.

\bigskip

\bigskip

Acknowledgment:  We wish to thank N. Bonesteel and J. Axe for calling
our attention to the ${1\over 8}$ effect and for helpful information
regarding it; A. Nersesayan and L. Gorkov for alerting us to the
literature on orbital antiferromagnetism [\oaf ];
A. H. Castro Neto for calling our attention to other
relevant work [\zaanen-\emerykiv]; and G. Baskaran, J. R.
Schrieffer and S. Treiman for several instructive discussions.

\endpage

\refout

\endpage

\figout

\endpage
\epsfysize=.8\hsize\hskip2cm\vskip3cm
\hbox to .8\hsize{\epsffile{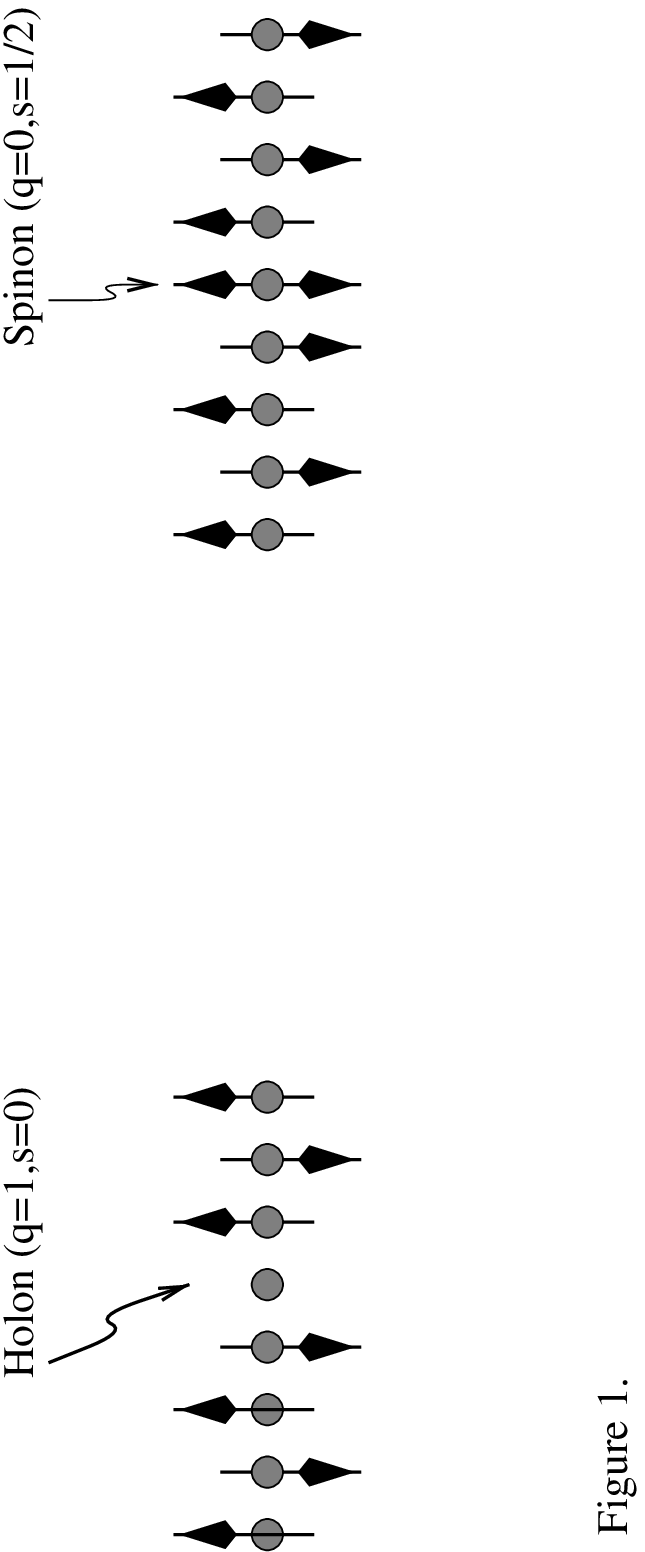}}\nextline\hskip-1cm
\endpage
\epsfysize=.8\vsize\hskip2cm\vskip1cm
\vbox to .8\vsize{\epsffile{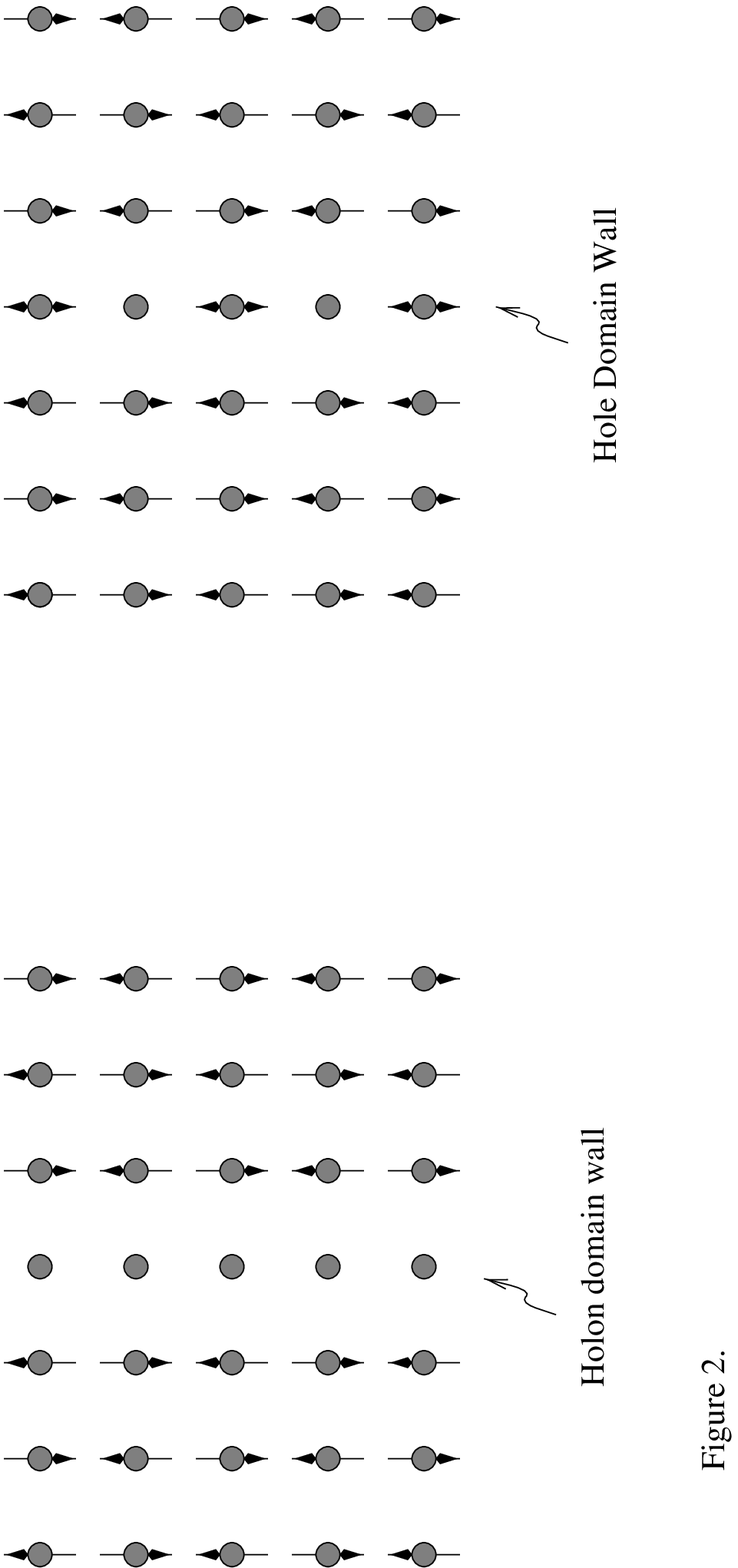}}\nextline\hskip-1cm
\endpage
\epsfysize=.8\vsize\hskip2cm\vskip1cm
\vbox to .8\vsize{\epsffile{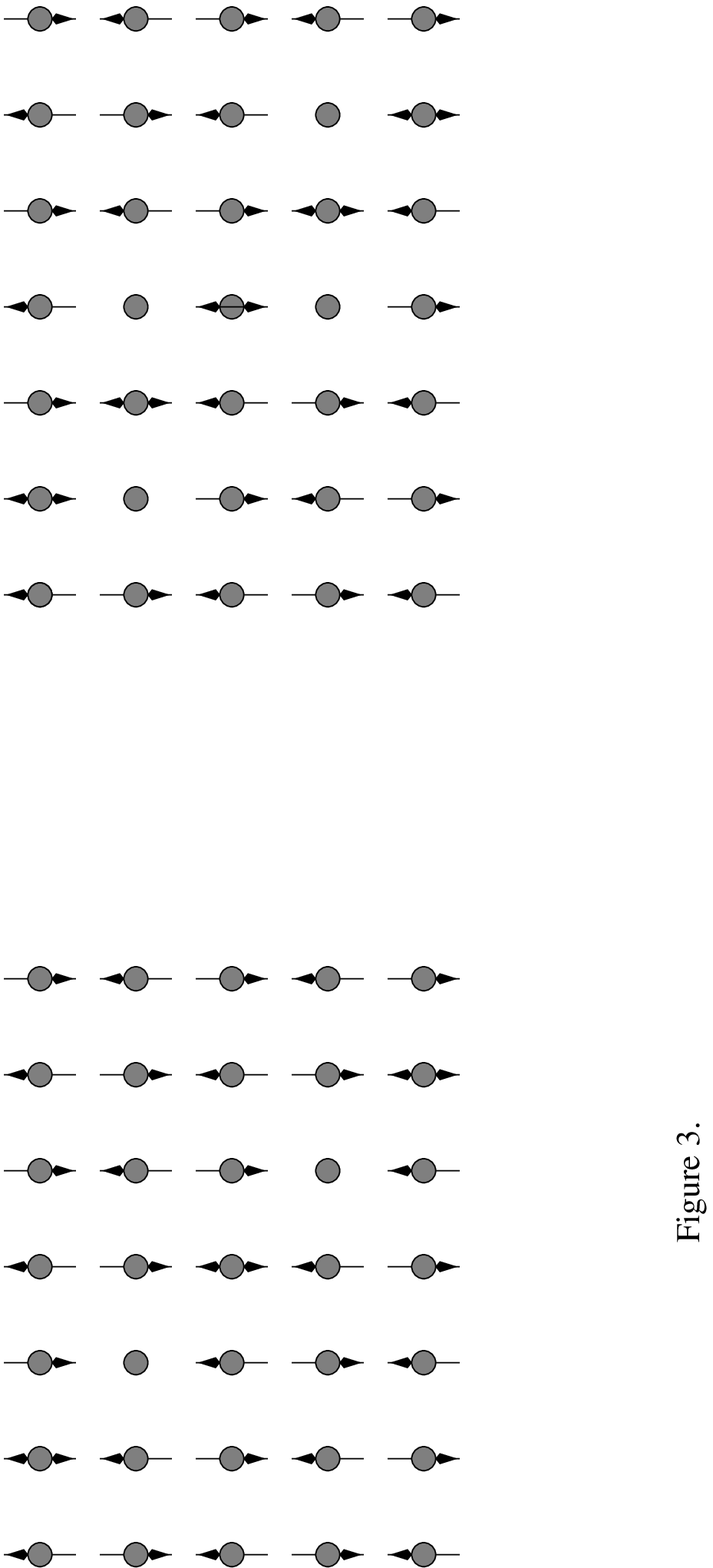}}\nextline\hskip-1cm
\endpage

\end